\documentclass[journal]{IEEEtran}
\usepackage{ucs}
\usepackage[utf8x]{inputenc}
\usepackage[cmex10]{amsmath}
\usepackage{cite, amsfonts, amssymb, amsthm, bm, bbm, graphicx, relsize, multirow, booktabs, tikz,subfigure,soul}
\usepackage[american]{babel}
\usepackage[T1]{fontenc}
\usepackage{algorithmic, algorithm}
\usepackage[multiple]{footmisc}
\theoremstyle{definition}

\theoremstyle{remark}

\renewcommand{\IEEEQED}{\IEEEQEDopen}

\begin{document}

\title{Subspace Tracking Algorithms for Millimeter Wave MIMO Channel Estimation with Hybrid Beamforming}
\author{Stefano Buzzi, {\em Senior Member}, {\em IEEE}, and Carmen D'Andrea
\thanks{The authors are with the Department of Electrical and Information Engineering, University of Cassino and Lazio Meridionale, I-03043 Cassino, Italy (buzzi@unicas.it, carmen.dandrea@unicas.it).}
}
\maketitle
\pagestyle{empty}
\begin{abstract}
This paper proposes the use of subspace tracking algorithms for performing MIMO channel estimation at  millimeter wave (mmWave) frequencies. Using a subspace approach, we develop a protocol enabling the estimation of the right (resp. left) singular vectors at the transmitter (resp. receiver) side; then, we adapt the projection approximation subspace tracking with deflation (PASTd) and the orthogonal Oja (OOJA) algorithms to our framework and obtain two channel estimation algorithms. The hybrid analog/digital nature of the beamformer is also explicitly taken into account at the algorithm design stage. Numerical results show that the proposed estimation algorithms are effective, and that they perform better than two relevant competing alternatives available in the open literature.
\end{abstract}


\section{Introduction}
The use of frequency bands in the range 10−100 GHz, a.k.a. millimeter waves (mmWaves), for cellular communications, is among the most striking technological innovations brought by fifth generation (5G) wireless networks \cite{whatwillbe}. 
Indeed, the scarcity of available frequency bands in the sub-6 GHz spectrum has been the main thrust for considering the use of higher frequencies for cellular applications, and indeed recent research \cite{itwillwork} has shown that mmWaves, despite increased path-loss and atmospheric absorption phenomena, can be actually used for cellular communications over short-range distances (up to 100-200 meters), provided that multiple antennas are used at both sides of the communication link: MIMO processing, thus, is one distinguishing and key feature of mmWave systems.
When considering mmWave links, one challenging task is that of MIMO channel estimation. This paper tackles this issue, and in particular the contribution of this paper can be summarized as follows.
\begin{enumerate}
\item
Using a subspace approach, we develop a protocol enabling the estimation of the right (resp. left) singular vectors at the transmitter (resp. receiver) side; then, we adapt the  PASTd algorithm  \cite{yang1995projection}, and the orthogonal Oja (OOJA) algorithm \cite{abed2000orthogonal} to our framework and obtain two subspace-based channel estimation algorithms.
\item
We adapt the proposed algorithms in order to take into account the hybrid analog/digital beamforming structure usually employed in mmWave wireless links. In particular, we assume that both at the transmitter and at the receiver the front-end RF modules are made by an analog  combining matrix whose columns correspond to beam-steering array response vectors, and show that the proposed channel estimation algorithms based on subspace tracking can be applied in this scenario too. 
\item
We compare the proposed algorithms with two competing alternatives available in the open literature, namely the   \textit{Approximate Maximum Likelihood} (AML) algorithm in \cite{haghighatshoar2016low,haghighatshoar2016massive}, and the \textit{Subspace Estimation using Arnoldi iteration} (SE-ARN) in \cite{ghauch2016subspace}. In this context, we also generalize the AML algorithm -- presented in \cite{haghighatshoar2016low,haghighatshoar2016massive} for the case in which the mobile station (MS) has just one antenna -- to the case in which the MS is equipped with multiple antennas. 
\item
We also show how the proposed channel estimation schemes can be used to implement a simple pilot-less differential modulation scheme in the case of multiplexing order one, and the corresponding symbol error probability (SER) is numerically evaluated. 
\end{enumerate}

This paper is organized as follows. Next Section contains the description of the system model, while in Section III the protocol to enable estimation of the left and right dominant singular vectors of the channel matrix using the subspace tracking algorithms is described. Section IV is devoted to the description of other estimation schemes, including the extension of the AML algorithm \cite{haghighatshoar2016low} to the case in which the MS is equipped with multiple antennas, and the SE-ARN.  Section V is devoted to the illustration of the numerical results, while, finally, concluding remarks are given in Section VI.

\textit{Notation:} We denote vectors by boldface lowercase letters (e.g. $\mathbf{x}$), matrices by boldface capital letters (e.g. $\mathbf{X}$), scalar constant by nonboldface letters (e.g $x$ or $X$). The superscripts $(\cdot)^T$ and $(\cdot)^H$ denote transpose and conjugate-transpose, respectively and the identity matrix of order $k$ is expressed with $\mathbf{I}_k$. 

\begin{center}
\begin{figure*}[t]
\includegraphics[scale=0.28]{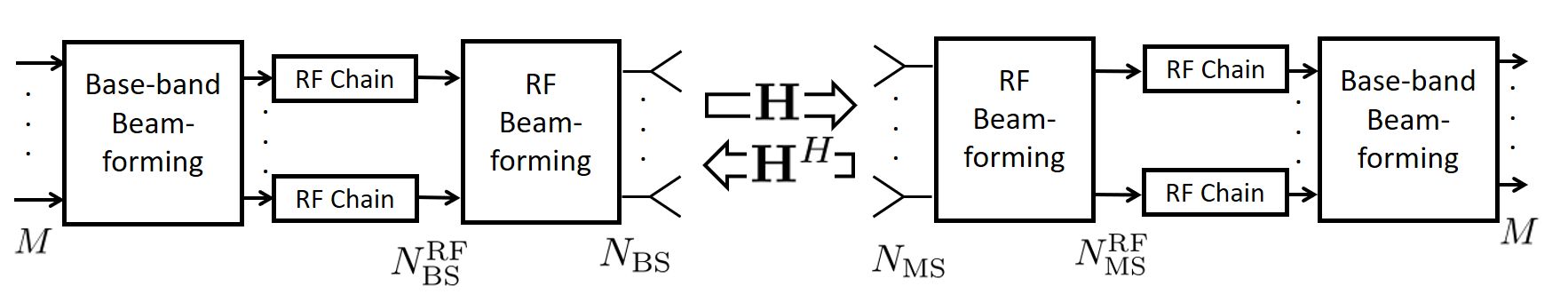}
\caption{Block scheme of the considered transceiver architecture.}
\label{fig:scheme}
\end{figure*}
\end{center}

\section{System model}

We consider a single-cell and single-user MIMO link that can be representative of the Base Station (BS) - Mobile Station (MS)  link in a wireless cellular system using an orthogonal multiple access scheme. The number of parallel streams, i.e. the multiplexing order, is denoted by $M$.  We consider a time-division-duplex (TDD) scenario, so that the BS-to-MS channel is the conjugate transpose of the MS-to-BS channel, provided that the transmission time does not exceed the channel coherence interval.
We denote by $N_{\rm BS}$ the number of antennas at the BS, and by $N_{\rm MS}$ the number of antennas at the MS, and assume for simplicity a bi-dimensional model, i.e. both the BS and the MS are equipped with a uniform linear array (ULA).
A schematic representation of the considered system is reported in Fig. \ref{fig:scheme}. In particular, beamforming at both sides of the link is of the hybrid type, in the sense that, in order to lower hardware complexity, analog signals combining is performed at RF in the transceiver front-end in order to reduce the dimensionality of the signals, and, then fully digital baseband combining is performed. We will denote by 
$N_{\rm BS}^{\rm RF}<N_{\rm BS}$ and $N_{\rm MS}^{\rm RF}<N_{\rm MS}$  the number of RF chains at the BS and at the MS, respectively.

\subsection{The clustered channel model}

We denote by $\mathbf{H}$ the $(N_{\rm MS} \times N_{\rm BS})$-dimensional matrix representing the BS-to-MS channel. Due to  TDD operation the reverse-link propagation channel is expressed as $\mathbf{H}^H$.
According to the popular narrowband clustered mmWave channel model 
\cite{spatiallysparse_heath,cairewsa2016,lee2014exploiting}, the channel  matrix is expressed as:
\begin{equation}
\mathbf{H}=\gamma\sum_{i=1}^{N_{\rm cl}}\sum_{l=1}^{N_{{\rm ray},i}}\alpha_{i,l}
\sqrt{L(r_{i,l})} \mathbf{a}_{\rm MS}(\phi_{i,l}^{\rm MS}) \mathbf{a}_{\rm BS}^H(\phi_{i,l}^{\rm BS}) + \mathbf{H}_{\rm LOS}\; .
\label{eq:channel1}
\end{equation}
In Eq. \eqref{eq:channel1}, 
we are implicitly assuming that the propagation environment is made of $N_{\rm cl}$ scattering clusters, each of which contributes with $N_{{\rm ray}, i}$ propagation paths, $i=1, \ldots, N_{\rm cl}$, plus a  possibly present LOS component.  
We denote by  $\phi_{i,l}^{\rm MS}$ and $\phi_{i,l}^{\rm BS}$ the downlink angle of arrival  (that coincides with the uplink angle of departure) at the MS and the downlink angle of departure (that coincides with the uplink angle of departure) at the BS of the $l^{th}$ ray in the $i^{th}$ scattering cluster, respectively. 
The quantities $\alpha_{i,l}$ and $L(r_{i,l})$ are the complex path gain and the attenuation associated  to the $(i,l)$-th propagation path. 
The complex gain  $\alpha_{i,l}\thicksim \mathcal{CN}(0, \sigma_{\alpha,i}^2)$, with  $\sigma_{\alpha,i}^2=1$  \cite{spatiallysparse_heath}. The factors $\mathbf{a}_{\rm MS}(\phi_{i,l}^{\rm MS})$ and $\mathbf{a}_{\rm BS}(\phi_{i,l}^{\rm BS})$ represent the normalized receive and transmit array response vectors evaluated at the corresponding angles of arrival and departure; for a ULA with half-wavelength inter-element spacing we have
$
\mathbf{a}_{\rm BS}(\phi_{i,l}^{\rm BS})=\displaystyle \frac{1}{\sqrt{N_{\rm BS}}}[1 \; e^{-j\pi \sin \phi_{i,l}^{\rm BS}} \; \ldots \; e^{-j\pi (N_{\rm BS}-1) \sin \phi_{i,l}^{\rm BS}}]^T$.  A similar expression can be also given for $\mathbf{a}_{\rm MS}(\phi_{i,l}^{\rm MS})$.
Finally, $\gamma=\displaystyle\sqrt{\frac{N_{\rm BS} N_{\rm MS}}{\sum_{i=1}^{N_{\rm cl}}N_{{\rm ray},i}}}$  is a normalization factor ensuring that the received signal power scales linearly with the product $N_{\rm BS} N_{\rm MS}$.
Regarding the LOS component, denoting by 
$\phi_{\rm LOS}^{\rm MS}$,  $\phi_{\rm LOS}^{\rm BS}$,
 the downlink arrival and departure angles corresponding to the LOS link, we assume that
\begin{equation}
\begin{array}{llll}
\mathbf{H}_{\rm LOS} = &  
I_{\rm LOS}(d) \sqrt{N_{\rm MS} N_{\rm BS} L(d)} e^{j \theta} \mathbf{a}_{\rm MS}(\phi_{\rm LOS}^{\rm MS})  \mathbf{a}_{\rm BS}^H(\phi_{\rm LOS}^{\rm BS}) \; .
\end{array}
\label{eq:Hlos}
\end{equation}
In the above equation, $\theta \thicksim \mathcal{U}(0 ,2 \pi)$, $d$ is the link length, while $I_{\rm LOS}(d) $ is a random variate indicating if a LOS link exists between transmitter and receiver, with $p$ the probability that $I_{\rm LOS}(d) =1$. 
A detailed description of all the parameters needed for the generation of sample realizations for the channel model of Eq. \eqref{eq:channel1} is reported in \cite{buzzidandreachannel_model}, and we refer the reader to this reference for further details on the channel model.

\section{Subspace-based channel eigendirections estimation}
Given the structured (parametric) channel model in \eqref{eq:channel1}, differently from what usually happens for MIMO channels at conventional sub-6 GHz frequencies where all the entries of the channel matrix are to be estimated, here we are actually interested only to the dominant left and right singular vectors of the channel matrix itself; it is also quite easy to realize that these singular vectors tend to coincide, in the limit of large number of antennas, with the ULA array responses at the angles corresponding to the rays with the complex gain with the largest norm.

\subsection{Subspace-based  channel estimation with fully-digital (FD) beamforming architecture}
We start by considering the case of FD beamforming, i.e. no analog beamforming is performed and the number of RF chains coincides with the number of antennas, both at the BS and at the MS. 
The proposed protocol for channel estimation consists of two successive phases. In  phase (a), the BS transmits a suitable probing signal and the MS estimates the dominant left eigenvectors of the channel matrix; then, in phase (b), the MS transmits a suitable signal and the BS estimates the dominant right eigenvectors of the channel matrix. 

With regard to phase (a), let  $\mathbf{s}_{\rm BS}(n)$, with $n=1, \ldots, P_{\rm BS}$, be a sequence of $N_{\rm BS}$-dimensional random column 
vectors with identity covariance matrix\footnote{As an example, a sequence of random uniform binary-valued antipodal symbols can be used.}. These vectors  are transmitted by the BS at (discrete) time $n=1, \ldots, P_{\rm BS}$; the signal received at the MS at time $n$ is expressed as
the following $(N_{\rm MS} \times 1)$-dimensional vector:

\begin{equation}
\mathbf{r}_{\rm MS}(n)=\mathbf{H}\mathbf{s}_{\rm BS}(n)+\mathbf{w}_{\rm MS}(n) \; ,
\label{eq:receivedMS}
\end{equation} 
where $\mathbf{w}_{\rm MS}(n)$ is the $N_{\rm MS}$-dimensional AWGN vector, modeled as $\mathcal{CN} (0, \sigma^2_n)$ independent RVs.
Letting  $\mathbf{H}=\mathbf{U}\mathbf{\Lambda}\mathbf{V}^H$ denote the singular value decomposition of the channel matrix,  the covariance matrix of the received signal is expressed as
\begin{equation}
\mathbf{R}_{\rm MS}=E\left[\mathbf{r}_{\rm MS}(n)\mathbf{r}_{\rm MS}^H(n)\right]=\mathbf{U}\mathbf{\Lambda}^2\mathbf{U}^H+ \sigma^2_n \mathbf{I}_{N_{\rm MS}} \; .
\label{eq:covMS}
\end{equation}
Given \eqref{eq:covMS}, it is thus easily seen that we can estimate the $M$ dominant left singular vectors of the channel matrix by estimating the $M$ dominant directions of the subspace spanned by the received vectors 
$\mathbf{r}_{\rm MS}(n)$, with $n=1, \ldots, P_{\rm BS}$. The signal processing literature is rich of adaptive subspace tracking algorithms that can be straightforwardly applied at the MS to obtain an estimate of the principal left singular vectors of the channel matrix. Deferring later the specification of the adopted subspace tracking algorithms,  let 
$\mathbf{D}_{MS}$ be the $(N_{\rm MS} \times M)$-dimensional matrix containing the estimate of the $M$ dominant singular vectors of $\mathbf{R}_{\rm MS}$; the matrix $\mathbf{D}_{\rm MS}$ will be used at the MS as a precoder during data transmission and as a combiner when receiving data from the BS. 
After $P_{\rm BS}$ symbol intervals, phase (a) is over and phase (b) starts. The MS now transmits random independent vectors with identity covariance matrix in order to enable estimation at the BS of the dominant right eigenvectors of the channel matrix $\mathbf{H}$. More precisely, 
let $\mathbf{s}_{\rm MS}(n)$, with $n=P_{\rm BS}+1, \ldots, P_{\rm BS}+ P_{\rm MS}$, be a sequence of 
$M$-dimensional random vectors with identity covariance matrix and assume that the MS transmits  the following $N_{\rm MS}$-dimensional data vectors
\begin{equation}
\mathbf{x}_{\rm MS}(n)=\mathbf{D}_{ \rm MS} \mathbf{s}_{\rm MS}(n),
\end{equation}
where $\mathbf{s}_{\rm MS}(n)$ is the $(M \times 1)$-dimensional vector whose entries take value in $\{\pm 1\}$.
The received discrete-time signal at the BS is represented by the following $N_{\rm BS}$-dimensional vector:
\begin{equation}
\mathbf{r}_{\rm BS}(n)=\mathbf{H}^H\mathbf{x}_{\rm MS}(n)+\mathbf{w}_{\rm BS}(n) \; ,
\label{eq:receiveBS}
\end{equation} 
where $\mathbf{w}_{\rm BS}(n)$ is the $N_{\rm BS}$-dimensional AWGN vector, modeled as $\mathcal{CN} (0, \sigma^2_n)$ independent RVs\footnote{Note that in phase (b) the MS is already using the estimated precoder
$\mathbf{D}_{ \rm MS}$; the illustrated procedure also works if the MS does not use the precoder and sends $N_{\rm MS}$-dimensional random vectors.}. 
Similarly to phase (a), under the assumption of negligible errors  in the estimation of the left singular vectors of the channel matrix, the covariance matrix of the received signal at the BS is expressed as
\begin{equation}
\mathbf{R}_{\rm BS} = 
E\left[\mathbf{r}_{\rm BS}(n)\mathbf{r}_{\rm BS}^H(n)\right] \approx
\mathbf{V}\mathbf{\Lambda}^2\mathbf{V}^H + \sigma^2_n \mathbf{I}_{N_{\rm BS}} \; ,
\end{equation}
thus implying that the 
 the $M$ dominant right singular vectors of the channel matrix can be estimated by running adaptive subspace tracking algorithms at the BS. We will denote by $\mathbf{D}_{\rm BS}$ is the $(N_{\rm BS} \times M)$-dimensional matrix containing the estimates of the $M$ dominant singular vectors of $\mathbf{R}_{\rm MS}$.

\subsection{Subspace-based  channel estimation with hybrid (HY) beamforming architecture}
In the previous section a FD beamforming structure has been assumed. We now examine the case in which, for complexity reasons, a HY beamforming architecture is adopted. The front-end processing consists of an analog RF combining matrix aimed at reducing the number of RF chains needed to implement the base-band processing. From a mathematical point of view, 
the beamforming matrices at the MS and at the BS can be expressed as
\begin{equation}
\begin{array}{llll}
\mathbf{D}_{\rm MS}&= & \mathbf{D}_{\rm MS,RF}\mathbf{D}_{\rm MS,BB} \; , \qquad \mbox{and} \\
\mathbf{D}_{\rm BS}&= & \mathbf{D}_{\rm BS,RF}\mathbf{D}_{\rm BS,BB} \; ,
\end{array}
\label{eq:HYcombiners}
\end{equation}
respectively. In \eqref{eq:HYcombiners}, 
 $\mathbf{D}_{\rm MS,RF}$ is an $(N_{\rm MS} \times N_{\rm MS}^{\rm RF})$-dimensional matrix with unit-norm entries, while  $\mathbf{D}_{\rm MS,BB}$ is an $(N_{\rm MS}^{\rm RF} \times M)$-dimensional matrix with no constraint on its entries. Similarly,  $\mathbf{D}_{\rm BS,RF}$ is an $(N_{\rm BS} \times N_{\rm BS}^{\rm RF})$-dimensional matrix with unit-norm entries, and $\mathbf{D}_{\rm BS,BB}$ is an $(N_{\rm BS}^{\rm RF} \times M)$-dimensional baseband combining matrix. 
 The design of HY analog/digital combiners is a vastly explored research topic; most papers try to find the hybrid combiner that best approximates, according to some criterion, the optimal FD combiner. In this paper, we use a simpler and different approach. We assume that $\mathbf{D}_{\rm MS,RF}$ and $\mathbf{D}_{\rm BS,RF}$ have a fixed structure and in particular contain on their column the ULA array responses corresponding to a grid of discrete angles spanning the range $[-\pi/2, \pi/2]$. In particular, letting 
 \begin{equation}
 \begin{array}{lll}
\theta_{\rm MS}(i)= \left(-\frac{\pi}{2}+\frac{\pi(i-1)}{N_{\rm MS}^{\rm RF}}\right)\; , \; \; i=1, \ldots, N_{\rm MS}^{\rm RF} \; , \\
\theta_{\rm BS}(i)= \left(-\frac{\pi}{2}+\frac{\pi(i-1)}{N_{\rm BS}^{\rm RF}}\right)\; , \; \; i=1, \ldots, N_{\rm BS}^{\rm RF} \; ,
 \end{array}
 \end{equation}
 the RF combiners have the following structure:
\begin{equation}
\begin{array}{lll}
\mathbf{D}_{\rm BS,RF}& =& \left[\mathbf{a}_{\rm BS}\left(\theta_{\rm BS}(1) \right), \ldots \mathbf{a}_{\rm BS}\left(\theta_{\rm BS}(N_{\rm BS}^{\rm RF}\right) \right] \; , \\
\mathbf{D}_{\rm MS,RF}& =& \left[\mathbf{a}_{\rm MS}\left(\theta_{\rm MS}(1) \right), \ldots \mathbf{a}_{\rm MS}\left(\theta_{\rm MS}(N_{\rm MS}^{\rm RF}\right) \right] \; .
\end{array}
\end{equation}
Now, focus on the scheme in Fig. \ref{fig:scheme} and consider the  cascade of the BS analog beamformer, the channel $\mathbf{H}$ and the MS analog beamformer; it is straightforward to show that  this cascade can be modeled through the matrix 
$\widetilde{\mathbf{H}}=\mathbf{D}^H_{\rm MS,RF} \mathbf{H} \mathbf{D}_{\rm BS,RF}$, of dimension 
$N_{\rm MS}^{\rm RF} \times N_{\rm BS}^{\rm RF}$. As a consequence, the channel estimation scheme outlined in the previous section under the assumption of FD beamforming can be now re-applied on the (reduced-dimension) composite channel $\widetilde{\mathbf{H}}$.

\subsection{PASTd algorithm}
The PASTd algorithm was introduced in \cite{yang1995projection}; one of its most popular applications is reported in the highly-cited paper \cite{wang1998blind}. In order to illustrate this algorithm, let $\mathbf{r}$ be an $(N \times 1)$-dimensional random vector with autocorrelation matrix $\mathbf{C}=E\left[\mathbf{r}\mathbf{r}^H\right]$. Consider the scalar function 
\begin{equation}
\begin{array}{lll}
J\left(\mathbf{W}\right)& =E\left[\|\mathbf{r}-\mathbf{W}\mathbf{W}^T\mathbf{r}\|^2\right] \\ &=\text{tr}(\mathbf{C})-2\text{tr}(\mathbf{W}^T\mathbf{C}\mathbf{W})+\text{tr}(\mathbf{W}^T\mathbf{C}\mathbf{W}\mathbf{W}^T\mathbf{W}),
\end{array}
\label{cost_function}
\end{equation}
with a $(N \times M)$-dimensional matrix argument $\mathbf{W}$, with $M<N$. It is shown in \cite{yang1995projection} that
\begin{itemize}
\item[-] the matrix $\mathbf{W}$ is a stationary point of $J\left(\mathbf{W}\right)$ if and only if 
$\mathbf{W}=\mathbf{T}_M\mathbf{Q}$, where $\mathbf{T}_M$ is a $(N \times M)$-dimensional matrix contains any $M$ distinct eigenvectors of $\mathbf{C}$ and $\mathbf{Q}$ is any $(M \times M)$-dimensional unitary matrix.
\item[-] All stationary points of $J\left(\mathbf{W}\right)$ are saddle points except when $\mathbf{T}_M$ contains the $M$ dominant eigenvectors of $\mathbf{C}$. In that case, $J\left(\mathbf{W}\right)$ attains the global minimum.
\end{itemize}
Therefore, for $M=1$ the solution of minimizing $J\left(\mathbf{W}\right)$ is given by the most dominant eigenvector of $\mathbf{C}$. In practical applications, only sample vectors $\mathbf{r}(i)$ are available and so the statistical average in \eqref{cost_function} is replaced by an exponentially-windowed time-average, i.e.:
\begin{equation}
J\left[\mathbf{W}(n)\right]=\sum_{i=1}^n{\beta^{n-i}\|\mathbf{r}(i)-\mathbf{W}(n)\mathbf{W}(n)^T\mathbf{r}(i)\|^2},
\label{cost_sum}
\end{equation}
where $\beta$ is the forgetting factor. The key trick of the PASTd approach is to approximate $\mathbf{W}(n)^T\mathbf{r}(i)$ in \eqref{cost_sum}, the unknown projection of $\mathbf{r}(i)$ onto the columns of $\mathbf{W}(n)$, by $\mathbf{y}(i)=\mathbf{W}(i-1)^T\mathbf{r}(i)$, which can be calculated for $i=1,\ldots n$ at sampling time $n$. This results in a modified cost function 
\begin{equation}
\tilde{J}\left[\mathbf{W}(n)\right]=\sum_{i=1}^n{\beta^{n-i}\|\mathbf{r}(i)-\mathbf{W}(n)\mathbf{y}(n)\|^2},
\label{modified_cost_sum}
\end{equation}
The recursive least squares (RLS) algorithm can then be used to solve for $\mathbf{W}(n)$ that minimizes the exponentially weighted least square criterion \eqref{modified_cost_sum}. The PASTd algorithm for tracking the eigenvalues and eigenvectors of the signal subspace is based on deflaction technique and its basic idea is as follows. For $M=1$, by minimizing $\tilde{J}\left[\mathbf{W}(n)\right]$ in \eqref{modified_cost_sum} the most dominant eigenvector is updated; then, the contributon from this estimated eigenvector is removed from $\mathbf{r}(n)$ itself, and the second most dominant eigenvector can be now extracted from the data. This procedure is applied repeatedly until the $M$ dominant eigenvectors are sequentially estimated.  The complete PASTd procedure is reported in Algorithm \ref{PASTd}.
\textit{Computational Complexity}: The PASTd algorithm costs $4NM + O(N)$ flops per iteration.

\begin{algorithm}[!t]

\caption{The PASTd Algorithm}

\begin{algorithmic}[1]

\label{PASTd}

\STATE $\mathbf{x}_1(n)=\mathbf{r}(n)$

\FOR { $m=1:M$}

\STATE  {$y_m(n)=\mathbf{u}_m^H(n-1)\mathbf{x}_m(n)$}

\STATE {$\lambda_m(n)=\beta\lambda_m(n-1)+|y_m(n)|^2$}

\STATE {$\mathbf{u}_m(n)\! \! = \! \! \mathbf{u}_m(\!n-1\!)\!+\!\left[\mathbf{x}_m(n)\!-\!\mathbf{u}_m(n-1)y_m(n)\right]\frac{y_m(n)^*}{\lambda_m(n)}$}

\STATE $\mathbf{x}_{m+1}(n)=\mathbf{x}_m(n)-\mathbf{u}_m(n)y_m(n)$

\ENDFOR

\end{algorithmic}

\end{algorithm}

 \subsection{OOJA algorithm}
The OOJA algorithm was introduced in \cite{abed2000orthogonal}; it builds upon the minor subspace extraction algorithm of Oja in \cite{oja1992principal}. According to Oja's algorithm the iteration is:
\begin{equation}
\begin{array}{lll}
\mathbf{W}(n+1)& =\mathbf{W}(n)-\delta\left[\mathbf{r}(n)\mathbf{v}^T(n)-\mathbf{W}(n)\mathbf{v}(n)\mathbf{v}^T(n)\right]\\ &=\mathbf{W}(n)-\delta\mathbf{p}(n)\mathbf{v}^T(n),
\end{array}
\label{OJA}
\end{equation}
where $\mathbf{W}$ is a $(N \times M)$-dimensional matrix, with $M<N$, $\mathbf{v}(n)= \mathbf{W}^T(n)\mathbf{r}(n)$, $\mathbf{p}(n)=\mathbf{r}(n)-\mathbf{W}(n)\mathbf{v}(n)$ and $\delta$ is a learning parameter. The OOJA algorithm consists of  iteration \eqref{OJA}, proposed in \cite{oja1992principal}, plus the following orthogonalization step of the weight matrix:
\begin{equation}
\mathbf{W}(n+1)=\mathbf{W}(n+1)\left(\mathbf{W}^T(n+1)\mathbf{W}(n+1)\right)^{-1/2},
\label{OOJA}
\end{equation}
where $\left(\mathbf{W}^T(n+1)\mathbf{W}(n+1)\right)^{-1/2}$ denotes the inverse square root of  $\mathbf{W}^T(n+1)\mathbf{W}(n+1)$. In \cite{abed2000orthogonal} the authors write \eqref{OJA} as:
\begin{equation}
\begin{array}{lll}
\!\!\mathbf{W}(n+1)\!& =\!\left[\mathbf{W}(n)-\delta\mathbf{p}(n)\mathbf{v}^T(n)\right] \left(\mathbf{I}_M+\tau(n)\mathbf{v}(n)\mathbf{y}^T(n)\right] \\ & =\mathbf{W}(n)-\delta\bar{\mathbf{p}}(n)\mathbf{v}^T(n),
\end{array}\label{OOJA_final}
\end{equation}

where 
\begin{equation}
\tau(n)=\frac{1}{\|\mathbf{v}(n)\|^2}\left[\frac{1}{\sqrt{1+\delta^2\|\mathbf{p}(n)\|^2\|\mathbf{v}(n)\|^2}}-1\right]
\end{equation} 
and 
\begin{equation}
\bar{\mathbf{p}}(n)=-\frac{\tau(n)\mathbf{W}(n)\mathbf{v}(n)}{\delta}+\left[1+\tau(n)\|\mathbf{v}\|^2\right]\mathbf{p}(n).
\end{equation}
The complete OOJA procedure is reported in Algorithm \ref{OOJA_Algorithm}.
\textit{Computational Complexity}: The OOJA algorithm costs $3NM + O(N)$ flops per iteration.
\begin{algorithm}[!t]

\caption{The OOJA Algorithm}

\begin{algorithmic}[1]

\label{OOJA_Algorithm}

\STATE $\mathbf{v}(n)=\mathbf{W}^T(n)\mathbf{r}(n)$

\STATE  $\mathbf{z}(n)=\mathbf{W}(n)\mathbf{v}(n)$

\STATE  $\mathbf{p}(n)=\mathbf{r}(n)-\mathbf{z}(n)$

\STATE  $\varphi(n)=\frac{1}{\sqrt{1+\delta^2\|\mathbf{p}(n)\|^2\|\mathbf{v}(n)\|^2}}$

\STATE  $\tau(n)=\frac{\phi(n)-1}{\|\mathbf{v}(n)\|^2}$

\STATE  $\bar{\mathbf{p}}(n)=-\frac{\tau(n)\mathbf{z}(n)}{\delta}+\phi(n)\mathbf{p}(n)$

\STATE  $\mathbf{W}(n+1)=\mathbf{W}(n)-\delta\bar{\mathbf{p}}(n)\mathbf{v}^T(n)$

\end{algorithmic}

\end{algorithm}

\begin{figure}[t]
\centering
\includegraphics[scale=0.25]{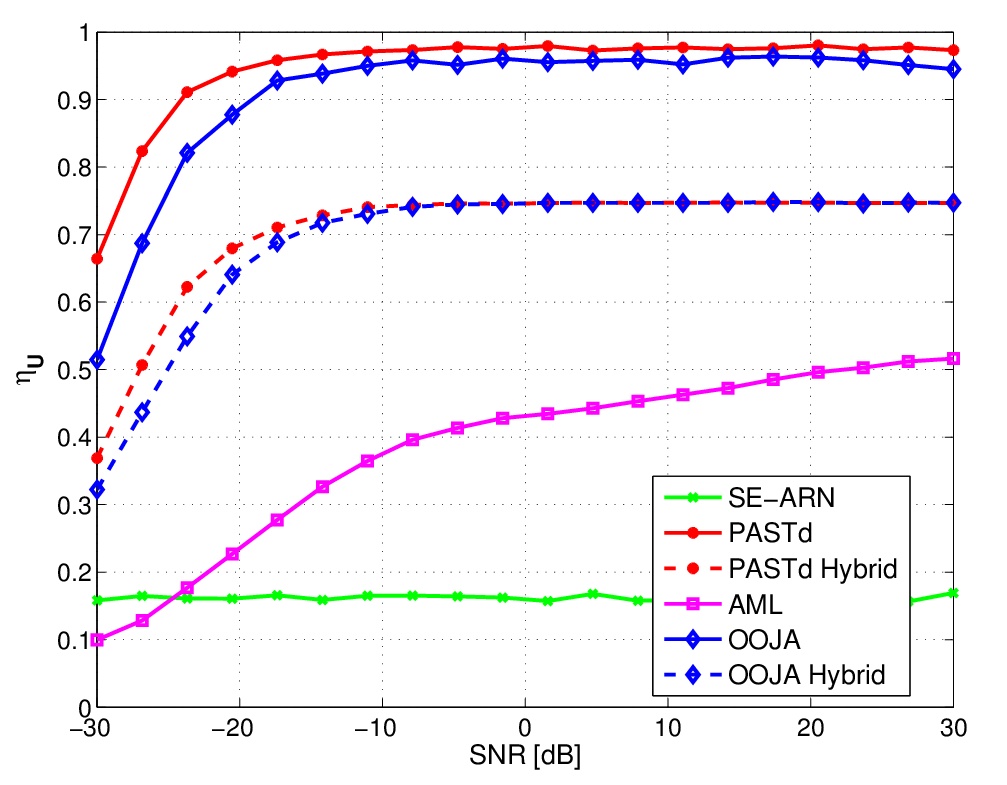}
\caption{$\eta_U$ versus the received SNR for a system with $N_{\rm MS} \times N_{\rm BS}=30 \times 100$. For the HY implementations we have used $N_{\rm MS}^{\rm RF}=10$ and  $N_{\rm BS}^{\rm RF}=20$.}
\label{Fig:fig_etaU_SNR}
\end{figure}

\begin{figure}[t]
\centering
\includegraphics[scale=0.25]{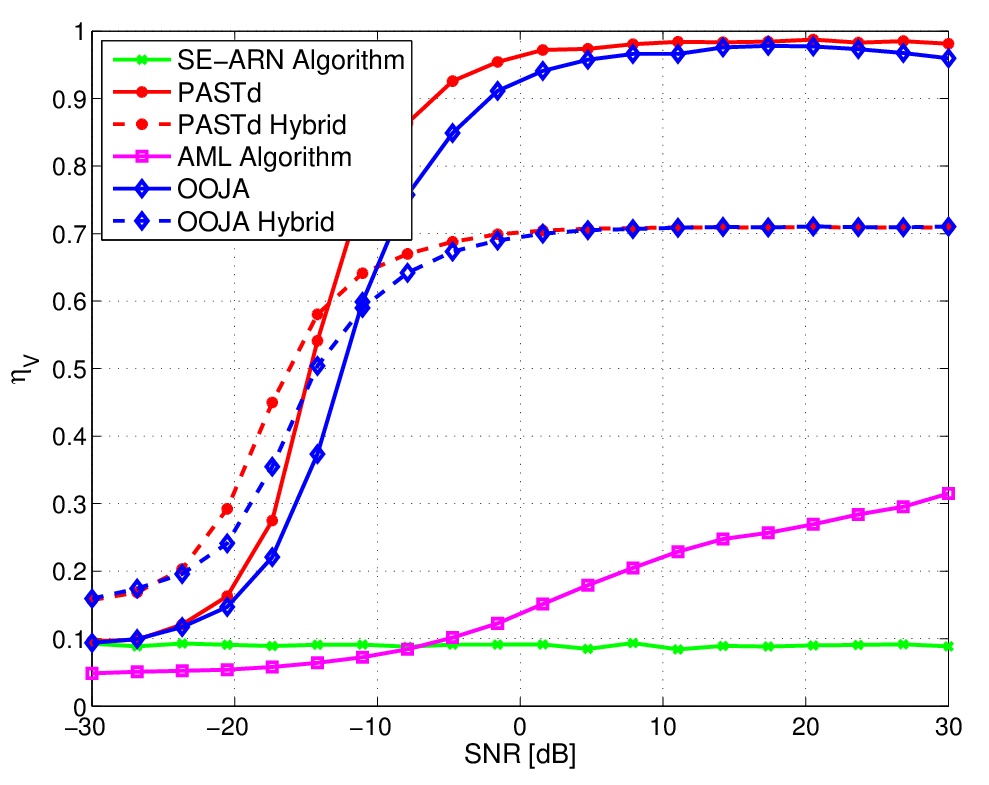}
\caption{$\eta_U$ versus the received SNR for a system with $N_{\rm MS} \times N_{\rm BS}=30 \times 100$. For the HY implementations we have used $N_{\rm MS}^{\rm RF}=10$ and  $N_{\rm BS}^{\rm RF}=20$.}
\label{Fig:fig_etaV_SNR}
\end{figure}
\section{Numerical Simulations}

We now provide some numerical results. The first performance measure that we are considering is the normalized correlation between the true channel eigenvectors and the estimated ones. In particular, we denote by $\mathbf{u}$ and  $\mathbf{v}$ the left and the right eigenvector of the channel matrix corresponding to the dominant eigenvalue respectively, and by $\mathbf{\hat{u}}$ and  $\mathbf{\hat{v}}$ the estimates of left and right dominant eigenvector of the channel matrix, respectively. The normalized correlations are defined as
\begin{equation}
\eta_U=\frac{\left|\mathbf{u}^H\mathbf{\hat{u}}\right|}{\|\mathbf{u}\| \|\mathbf{\hat{u}}\|},
\end{equation}
and as
\begin{equation}
\eta_V=\frac{\left|\mathbf{v}^H\mathbf{\hat{v}}\right|}{\|\mathbf{v}\| \|\mathbf{\hat{v}}\|}.
\end{equation}
We will also report the achievable rate and the error probability assuming differential phase shift keying signaling and differential non-coherent detection - notice indeed that the proposed channel estimation methods do not rely on known training symbols, and the dominant eigenvector is estimated with no information on the signal phase.
 
In our simulation setup, we consider a communication bandwidth of $W = 500$ MHz centered
over the carrier frequency $f_0=73$ GHz. The distance between the transmitter and the receiver is 50 m; the additive thermal noise is assumed to have a power spectral density of $-174$ dBm/Hz, while the front-end receiver at the BS and at the MS is assumed to have a noise figure of $3$ dB. The shown results come from an average over 500 random scenario realizations with independent channels. 
 
\begin{figure}[t]
\centering
\includegraphics[scale=0.25]{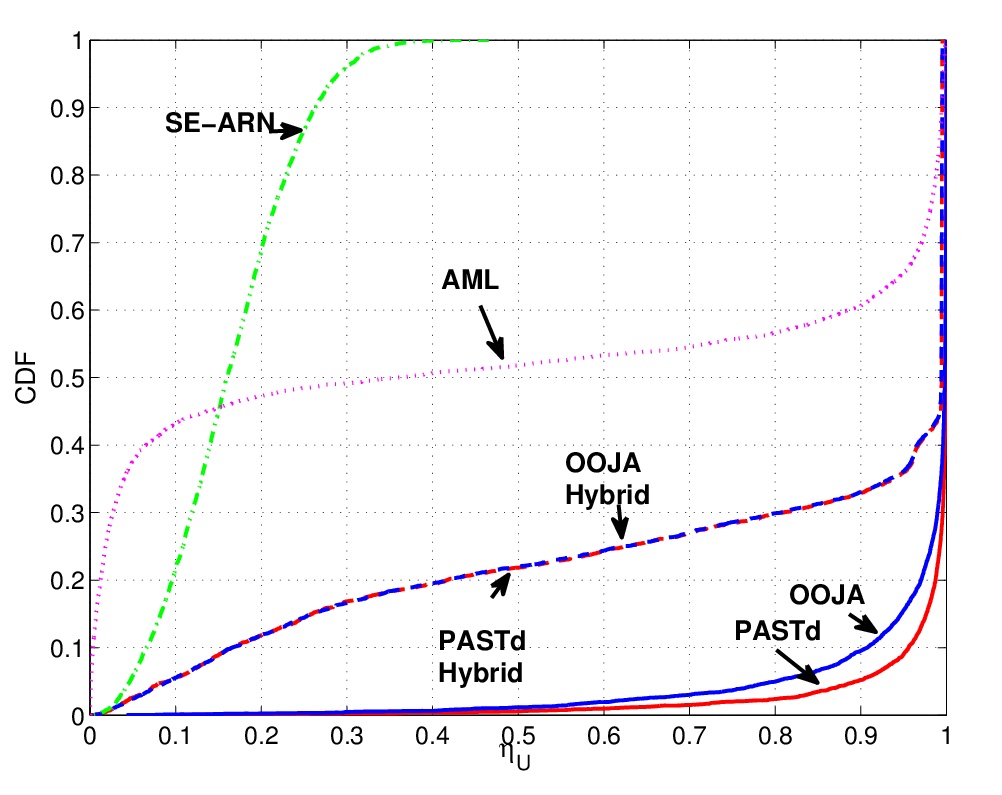}
\caption{CDF of $\eta_U$ for a system with $N_{\rm MS} \times N_{\rm BS}=30 \times 100$. For the HY implementations we have used $N_{\rm MS}^{\rm RF}=10$ and  $N_{\rm BS}^{\rm RF}=20$.}
\label{Fig:fig_etaU_CDF}
\end{figure}

\begin{figure}[t]
\centering
\includegraphics[scale=0.25]{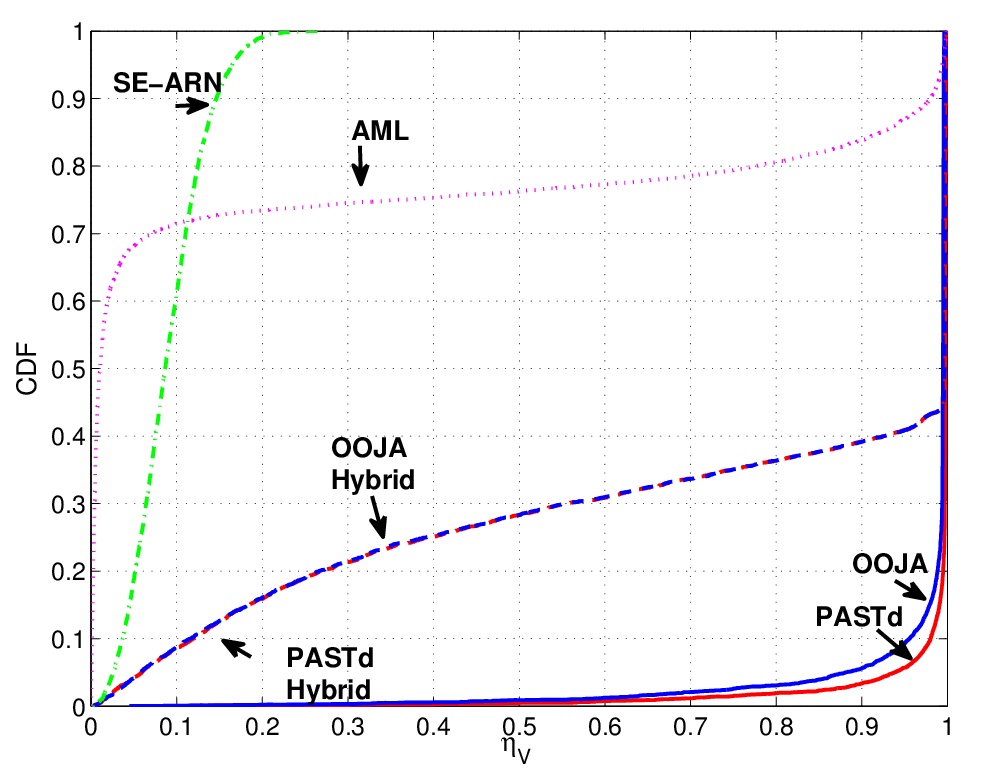}
\caption{CDF of $\eta_V$ for a system with $N_{\rm MS} \times N_{\rm BS}=30 \times 100$. For the HY implementations we have used $N_{\rm MS}^{\rm RF}=10$ and  $N_{\rm BS}^{\rm RF}=20$.}
\label{Fig:fig_etaV_CDF}
\end{figure}

\begin{figure}[t]
\centering
\includegraphics[scale=0.25]{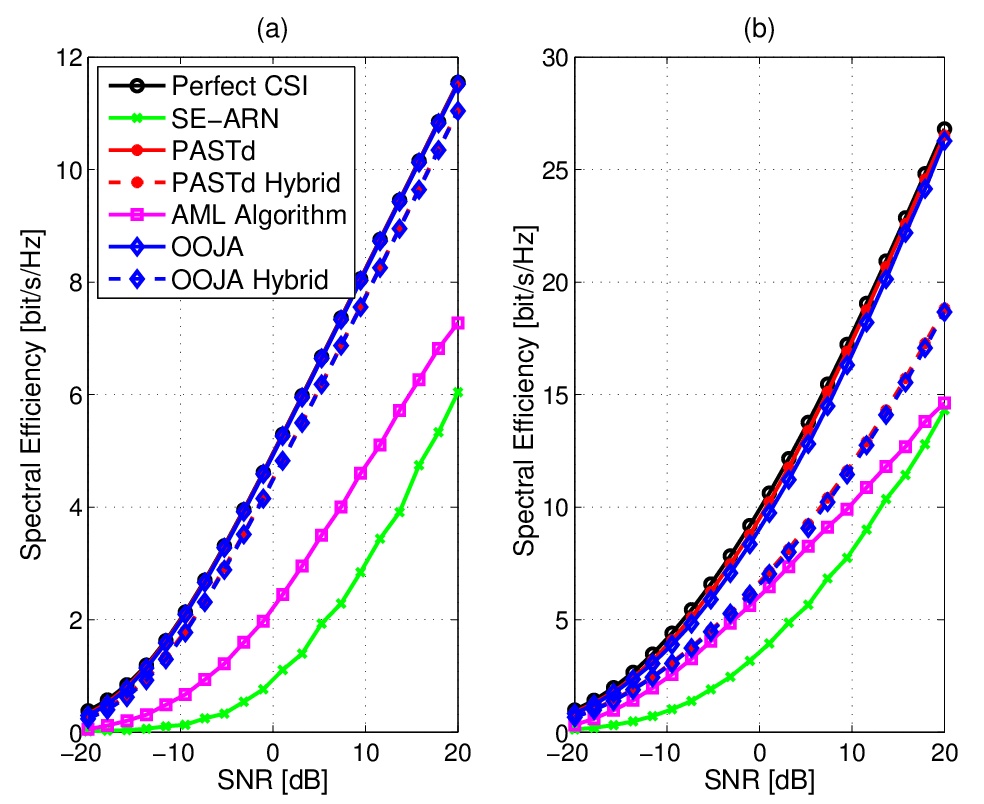}
\caption{Spectral efficiency for the BS-to-MS link versus received SNR. Parameters: $N_{\rm MS} \times N_{\rm BS}=30 \times 100$, $N_{\rm MS}^{\rm RF}=10$, $N_{\rm BS}^{\rm RF}=20$, multiplexing order $M=1$ in subplot $(a)$ and $M=3$ in subplot $(b)$.}
\label{Fig:fig_SE}
\end{figure}

\begin{figure}[t]
\centering
\includegraphics[scale=0.25]{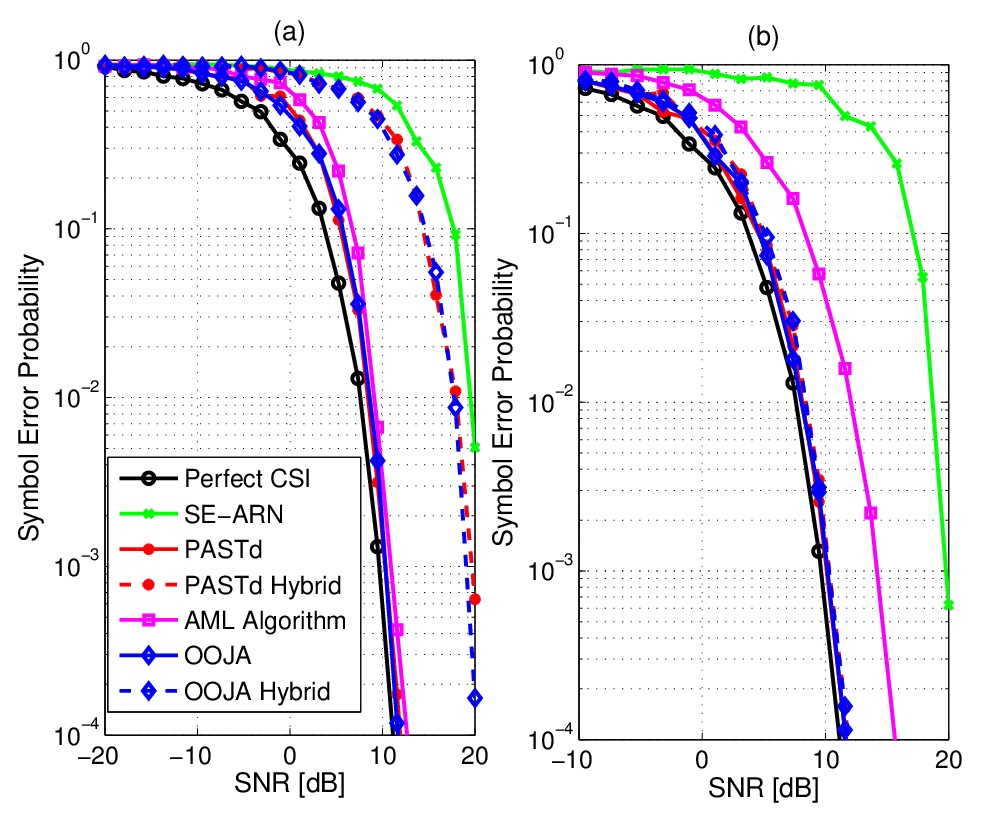}
\caption{Error probability for the BS-to-MS link versus received SNR. Parameters: 16-PSK differential modulation, $N_{\rm MS} \times N_{\rm BS}=30 \times 100$, $N_{\rm MS}^{\rm RF}=10$, $N_{\rm BS}^{\rm RF}=20$, 10 training symbols in subplot $(a)$, and 50 training symbols in subplot $(b)$.}
\label{Fig:fig_proberr}
\end{figure}

The length of the training phase for all the algorithms is $P_{\rm BS}=P_{\rm MS}=30$; of these training samples, the first ten are used to perform an SVD of the sample covariance matrix of the data, and the corresponding dominant eigenvector is used to initialize the PASTd and OOJA algorithms.  
In Figs. \ref{Fig:fig_etaU_SNR} and \ref{Fig:fig_etaV_SNR} we report  $\eta_U$ and $\eta_V$ versus the received SNR for the considered algorithms and also for the SE-ARN and AML algorithms. 
In Figs. \ref{Fig:fig_etaU_CDF} and \ref{Fig:fig_etaV_CDF} we report instead the cumulative distribution function (CDF) of $\eta_U$ and $\eta_V$  at an SNR of 10 dB, respectively  -- the remaining parameters are the same as the ones taken in Figs.  \ref{Fig:fig_etaU_SNR}-\ref{Fig:fig_etaV_SNR}. First of all, it is seen that the proposed subspace tracking algorithms achieve good performance and outperform competing alternatives. As expected, FD implementations achieves better performance than their HY suboptimal implementations. Further investigations are actually needed here to test the system performance with different types of analog front-end beamformers. 
Fig. \ref{Fig:fig_SE} reports the achievable spectral efficiency for the BS-to-MS link versus SNR, for multiplexing order $M=1$ (in subplot (a)) and $M=3$ (in subplot (b)). In particular, the plotted performance measure is the following
\begin{equation}
\begin{array}{lll}
\mathcal{R}_{\rm MS}=  \log_2 \det &\left[ \mathbf{I}_M + P_{\rm T, BS}\left(\sigma^2_n \mathbf{D}_{\rm MS}^H\mathbf{D}_{\rm MS} \right)^{-1} \; \times \right. \\ \\
&  \left. \mathbf{D}_{\rm MS}^H \mathbf{H}\mathbf{D}_{\rm BS}\mathbf{D}_{\rm BS}^H
\mathbf{H}^H\mathbf{D}_{\rm MS}\right] \; .
\end{array}
\label{eq:ASE}
\end{equation}
with $P_{\rm T, BS}$ the transmitted power at the BS. 
Results again show that the proposed algorithms have good performance levels. 
Finally, Fig. \ref{Fig:fig_proberr} we report the symbol error probability versus SNR for the BS-to-MS link and assuming multiplexing order $M=1$. Subplot (a) refers to the case in which the training phase length is 10 (with two symbols used to perform the SVD to initialize the PASTd and OOJA algorithms), while subplot (b) refers to the case in which the training phase length is 50 (with 10 symbols used for initialization). A differential 16-PSK modulation has been assumed. Overall, results confirm again the good performance of the proposed estimation techniques, even for the case of short training length.

%
%
%

\section{Conclusion}
This paper has been focused on the problem of channel estimation for wireless single-user MIMO links at mmWave frequencies. Exploiting the clustered propagation channel model, it is shown that subspace tracking algorithms can be used in order to track the principal eigenvectors of the channel matrix. Results have shown that the proposed estimation algorithms are effective and capable of attaining good performance levels also for very short training phases. This research can be extended in several directions. First of all, for the HY solutions, we have used a fixed analog beamformer. Further research is needed in order to properly design more effective analog beamformers, possibly to be tuned via an adaptive algorithm. Additionally, note that we have been proposing here a single-user estimation technique; multiuser joint  channel learning schemes are thus to be investigated. Finally, it would be of great interest to investigate the performance of the proposed adaptive algorithms in a dynamic environment, wherein channel impulse response changes due to, e.g., MS mobility.

\bibliographystyle{IEEEtran}

\bibliography{FracProg_SB,finalRefs,references}

\end{document}